
\vbadness=10000
\font\twelvebf=cmb10 scaled 1200
\font\twelvesl=cmsl10 scaled 1200
\hsize 14.5 truecm
\vsize 21 truecm
\catcode`\@=11
\textfont0=\tenrm \scriptfont0=\sevenrm \scriptscriptfont0=\fiverm
\def\rm{\fam\z@\tenrm}
\textfont1=\teni \scriptfont1=\seveni \scriptscriptfont1=\fivei
\def\mit{\fam\@ne} \def\oldstyle{\fam\@ne\teni}
\textfont2=\tensy \scriptfont2=\sevensy \scriptscriptfont2=\fivesy
\def\cal{\fam\tw@}
\def\sll{\twelvesl}
\def\etal{\hbox{\fit{et{}al.{ }}}}
\def\singlespace{\baselineskip=14pt}

\def\skuno{\vskip 20pt}
\def\bff{\twelvebf}
\def\blankline{\par\vskip 8pt}
\def\etal{\sll et al.}

\def\simless{\lapp}
\def\lapp{\mathbin{\raise2pt \hbox{$<$} \hskip-9pt \lower4pt \hbox{$\sim$}}}
\def\gapp{\mathbin{\raise2pt \hbox{$>$} \hskip-9pt \lower4pt \hbox{$\sim$}}}
\def\parref{\par\noindent\hangindent 1 cm}
\singlespace
\baselineskip 20pt
\centerline {\bff 1. Introduction}
\blankline\noindent
The role played by the intergalactic medium in determining the morphology and
evolution of radio sources has been considered in several papers (see e.g.
Fanti 1984, Burns 1986). The external gas can interact with a radio source in
different ways: confining the source, modifying its morphology via
ram-pressure and possibly feeding the active nucleus via accretion.

Previous studies at X-ray wavelengths (see Sarazin 1986 for a review) have
shown that the radio galaxies in clusters are embedded within a diffuse hot
gas at temperatures of $> 10^7$ K, with densities of a few $10^{-3} $cm$^{-3}$.
Furthermore, in many cases a hot corona is associated with the galaxy itself.

In particular, from the analysis of data from the Einstein data bank, it
appears
that the gas in rich clusters and the hot X-ray coronae of individual galaxies
may not be in pressure equilibrium with the relativistic particles in the
radio plasmoids of low luminosity radio galaxies (e.g. Morganti {\etal} 1988,
Feretti {\etal} 1990, 1992). A definite
answer to the confinement question is important in order to reconstruct
reliable models of radio source evolution.

Radio sources exhibiting a "head-tail" structure represent an important
example of the interaction between radio plasmoids and intergalactic medium.
These sources have been found to be associated both with galaxies in rich
clusters and galaxies in groups (Ekers {\etal} 1978, Burns {\etal} 1987),
indicating the presence of significant ambient gas in groups.
In this framework, we discuss here the radio
and X-ray properties of the tailed radio galaxies B2 1615+35 (NGC 6109) and B2
1621+38 (NGC 6137). They were found in a systematic search of bright galaxies
associated with radio sources from the B2 catalog (Colla {\etal} 1975).

Observations with the Westerbork Synthesis Radio Telescope (WSRT) (Ekers
{\etal}, 1978) revealed that both sources exibit a  distorted radio structure,
head-tail type, although on completely different linear scales. The source
1615+35 has a radio structure extending more than $10^{\prime}$ ($\gapp 0.5$
Mpc\footnote{$^{\ddag}$} {The Hubble constant H$_{\circ}$=50 km s$^{-1}$
Mpc$^{-1}$ is assumed throughout the paper}), while  1621+38 has a total
extent of $\approx 45^{\prime \prime}$ ($\approx 40$ kpc), i.e. smaller than
the optical galaxy dimension. Both sources were reobserved with the Very Large
Array (VLA) at 20 cm, as part of a bigger program of observation of all the B2
 low luminosity radio galaxies (Fanti {\etal} 1987, and references therein).

The radio source 1615+35 is associated with an elliptical galaxy of
$m_{pg}$=14.9
at a redshift of 0.0296 (distance $D=177$ Mpc). It is
$\approx 1^{\prime}$ extended, corresponding to a size of $\approx 50$ kpc,
and lies, in projection, in the middle of the Zwicky cluster 1615.8+3505 which
is "medium compact". The radio source 1621+38 is associated with the
elliptical galaxy NGC 6137, which has a magnitude $m_{pg}$=14.1, a size
of $\approx 1.9^{\prime} \times 1.2^{\prime}$ ($\approx 100 \times 60$ kpc)
and a redshift of 0.031 ($D=186$ Mpc). It belongs to a small concentration of
galaxies at one edge of the cluster Zw 1611.6+3717, which extends over a large
area of sky (Ulrich 1978) and is classified "open" by Zwicky and Herzog (1966).

These two groups of galaxies appear to belong to a large supercluster,
including the Abell clusters 2199, 2197 and 2162, which have almost identical
redshifts (Noonan, 1973) and extend over a region $\sim$60 Mpc in size.

{}From the WSRT observations (see Ekers {\etal} 1978) a third radio source with
extended structure was detected  in the field and identified with NGC 6107.
The radio source size is $\approx 0.9^{\prime}  \times 0.7^{\prime}$; the
galaxy has a magnitude  $m_{pg}$ = 14.7 and belongs to the same
group as 1615+35 (redshift = 0.0306, $D=184$ Mpc, which implies an optical
extent $\approx 50 \times 40$ kpc).

Previous X-ray observations of the 1615+35 field are reported in the Einstein
Observatory Catalog of IPC X-ray sources (Vol. 6, p. 58, Seq. number I-322,
Harris {\etal} 1990). An extended emission appears, but no detailed analysis
has been published.

The sources 1615+35 and 1621+38 have been targets of pointed observations at
soft X-ray wavelengths by the ROSAT satellite. The source 1615+35 was observed
with both the Position Sensitive Proportional Counter (PSPC) and High
Resolution Imager (HRI), to obtain images with high and moderate angular
resolution as well as useful spectral information. The source 1621+38 was
observed with the HRI to obtain images with an angular resolution comparable
to the radio structure. Here we present the results of these observations with
the following outline. In Sec. 2 we report the details of the observations
and data reduction. In Sec. 3 the morphological and spectral
analysis is presented, as well as the optical identification of the detected
sources. In section 4 we outline the physical parameters deduced from the
X-ray data and radio observations. The results, the confinement of radio
components and the astrophysical implications  are discussed in the final Sec.
5.

\blankline\noindent
\blankline\noindent
\centerline{\bff 2. Observations and data reduction}
\blankline\noindent
The observational details are summarized in Table 1. A description of the
instrument can be found in Tr\"umper (1983) and Pfeffermann {\etal}
(1987).

The PSPC observation of 1615+35 was performed in two exposures of almost equal
duration. The data from the two exposures were first analyzed separately for
check of consistency and then merged.

The data analysis has been performed with the EXSAS package (Zimmerman {\etal}
1993), using the January 93 release of the response matrix for the two
detectors. The maps
of the X-ray brightness distribution were smoothed with circular
gaussians.

The counts of each source have been computed by integration on a circle around
the source centroid up to a radius depending on the source extent,
and subtraction of  a background, selected in a source free
region. In the spectral analysis the energy channels 17 - 235 have been
used, corresponding to an enery range 0.2-2.1 keV.

The positional errors, as quoted from the standard analysis, are about
0.15$^s$ in RA and 1.5$^{\prime\prime}$ in DEC. The overlay of the HRI
X-ray field of 1615+35 onto the optical and radio images (for the
optical and radio data, see below) shows good agreement of the source
positions within the errors. For the 1621+38 field a significant
displacement, mostly in RA, is found between the position of the radio
and optical images (coincident within $\sim 1^{\prime \prime}$) and the
peak of the X-ray emission.  By comparing the X-ray position of the field
radio  source 1622+38 (Gregory \& Condon 1990) with its VLA  radio
position,  obtained by us to an accuracy of $<1^{\prime \prime}$, we
evaluated the systematic error to be -0.4$^s$ in RA and -1$^{\prime
\prime}$ in Dec.

Optical maps of the fields were obtained from the digitized Palomar Sky
Survey (PSS). For the optical identifications of the X-ray sources we used the
General Stellar Catalogue (GSC) of the Hubble Space Telescope (HST),
the Automated Plate Measuring (APM) survey of the PSS (McMahon 1991), and
the Nasa/IPAC Extragalactic Database.

The radio maps used here were obtained previously with the Westerbork
Synthesis Radio Telescope (WSRT) and the Very Large Array (VLA). Furthermore,
new data were collected in August 1993 with the VLA in C configuration
as follows: a) the field of 1615+35  was observed at 20 cm to get
radio information on the source NGC6107 (see Sects. 3.1 and 3.2); b) the field
of 1621+38 was observed at 3.6 cm, with two pointings, for the accurate
radio-X-ray overlay (see  above).

The VLA data were reduced with the AIPS package, following the standard
procedure (Fourier inversion, CLEAN and RESTORE). In Table 2 we summarize the
available radio maps, and their  parameters.

\blankline\noindent
\blankline\noindent
\centerline{\bff 3. Results}
\blankline\noindent
{\it 3.1. B2 1615+35 - HRI Image}
\par\noindent
The HRI image smoothed with a gaussian of $12^{\prime \prime}$
(FWHM) is shown in Fig. 1, overlayed on the optical plate.
Six strong sources are detected within 10$^{\prime}$ from the field
center. They are labelled A$_1$ and A$_2$ (separation $\approx 65^{\prime
\prime}$), B$_1$ and B$_2$ (separation $\approx  35^{\prime \prime}$), C and
D. All sources appear pointlike, except for A$_1$ which shows an extension
of $\approx 40^{\prime \prime}$. Source positions, count rates and
identifications are given in Table 3.

The source B$_1$ is identified with our target, the B2 radio galaxy 1615+35
(NGC 6109). The enlargement of the HRI map and the high resolution radio map of
this object are given in Fig. 2a and b, respectively. The X-ray emission of
1615+35 appears pointlike, i.e. its extent is $\lapp 4^{\prime \prime}$ which
corresponds to a linear dimension of $\lapp 4$
kpc. We note that the X-ray emission coincides with
the radio core, which is the westernmost of the two radio peaks. Therefore,
this emission is likely to originate from the galaxy nucleus.

The source A$_1$ (Fig. 3a) is
identified with the radio emitting elliptical galaxy NGC 6107 (see
Ekers {\etal} 1978). The radio structure at 1.4 GHz (Fig.
3b) is dominated by the core,
and shows two opposite short jets with a total extent of $\sim 1^{\prime}$.

A radial profile of the X-ray emission, obtained by integrating counts
over concentric annuli of $2.5^{\prime \prime}$ in
size, centered on the X-ray peak, shows that the X-ray
emission is resolved with an extent corresponding to the
approximate size of the optical galaxy. The contribution of a possible
pointlike component is  estimated from fits to be
less than 30\% of the total flux.
This observed profile was then fit with the surface brightness
profile expected from a hydrostatic isothermal model (Cavaliere and
Fusco-Femiano 1976, 1981; Sarazin 1986), i.e.
$$I(r)=I_0(1+r^2/r^2_c)^{-3\beta+0.5} \eqno(1) $$
\noindent where r$_c$ is the core radius, $\beta$ is the ratio of the
star to gas temperature and $I_0$ represents the central brightness,
to get estimates of the gas parameters (see Sect. 4.1).
The HRI point spread function was taken into account in the fit.

Source A$_2$ coincides with a star of $m_v=8.9$ (n. 1117/2583 of the G.S.C.).
The sources B$_2$, C  and D are identified with  bluish stellar objects fainter
than 19.5$^m$ on the blue PSS plate.

\blankline\noindent
{\it 3.2. B2 1615+35 - PSPC Image}
\par\noindent
The image obtained with the full energy band of PSPC, smoothed
with a gaussian of
50$^{\prime \prime}$  (FWHM)
is presented in Fig. 4a. The field is dominated by 5 discrete
sources, which correspond to the sources detected with the HRI,
with B$_1$+B$_2$ merged together because of the lower spatial resolution.
\noindent
The most interesting feature in this map is the diffuse emission permeating
the center, which is better visible in Fig. 4b, obtained with a smoothing
function which enhances the low brightness features. The diffuse source is
clearly associated with the ambient medium of the galaxy group. It appears
elongated in the N-S direction (size $\approx 20^{\prime} \times 15^{\prime}$,
which means projected dimensions $\approx 800 \times 600$ kpc). Its count
rate, obtained after subtraction of the background and of the embedded
discrete sources, is $0.43 \pm 0.02$ s$^{-1}$.

A radial profile of the X-ray brightness of the diffuse source was obtained by
integrating the counts over concentric annuli of $30^{\prime \prime}$ in size,
centered on the centroid of the diffuse X-ray emission (RA$_{2000}$ =
16$^h$17$^m$31.0$^s$, DEC$_{2000}$ = 34$^{\circ}$ 58$^{\prime}$ 4$^{\prime
\prime}$). The  best fit of the model (Eq. 1, where now $\beta$ is the
galaxy to gas temperature)
to the observed profile  (see Fig. 5) gives the following estimate of the
parameters, with a reduced
$\chi^2$=1.9: $\beta=0.56\pm$0.07, a core radius
$r_c = 375\pm$85 arcsec, and I$_0$=(6.6$\pm$0.6)$\times$ 10$^{-7}$ count
s$^{-1}$ arcsec$^{-2}$.
Apparently, the diffuse X-ray emission is rather far
from spherical symmmetry and this may prevent a better model fitting.
\blankline\noindent
{\it 3.3. B2 1615+35 - Spectral Analysis}
\par\noindent
The emission from the diffuse component detected in the PSPC was fit with
a hot, optically thin plasma model (Raymond and Smith 1977).
The parameters of
the fit are the hydrogen column density N$_H$, the gas temperature $kT$, the
relative abundances of heavy elements ${\cal {M}}$ (see Allen {\etal} 1973),
and the normalization amplitude $A$.
As a first step, we fixed the  metallicity to
half of the standard cosmic abundance (${\cal {M}} = 0.5 $), in agreement with
the available data on clusters of galaxies (Rothenflug \& Arnaud 1985).
In this way, we obtained N$_H = 1.05 \pm 0.4 \times 10^{20}$
cm$^{-2}$, $kT = 2.4 \pm 1.2$ keV with reduced $\chi^2 = 0.9$ for 35 d.o.f.
The value of the hydrogen column density is consistent with the expected
galactic absorption at the source position (N$_H = 1.52 \times 10^{20}$
cm$^{-2}$, Dickey \& Lockman 1990). Then we fixed N$_H$ to the galactic value,
and performed an analysis of the confidence levels of ${\cal {M}}$ vs $kT$
(see the $\chi^2$ contour plots in Fig. 6a). We found that the metallicity
is significantly lower than 0.5, with the most likely value ${\cal {M}}
\approx 0.1$. The best fit by fixing N$_H$ and ${\cal {M}}=0.1$ gives a
temperature $kT = 2.1 \pm 0.6$ keV and a normalization amplitude $A=(6.7 \pm
0.4) \times 10^{-3}$ counts s$^{-1}$ keV$^{-1}$, with $\chi^2$=0.8 (see Fig.
6b). With these parameters, 1 count s$^{-1}$ corresponds to an energy flux of
$1.2 \times 10^{-11}$ erg s$^{-1}$ cm$^{-2}$, and the total luminosity of the
diffuse source in the  ROSAT energy band (0.1-2.4 keV) is $L_X= 2 \times
10^{43} h^{-2}_{50}$ erg s$^{-1}$, where $h_{50}$= H$_0$/50
km s$^{-1}$ Mpc$^{-1}$.

As a further test, a power law spectrum has been fit to the photon flux, and
an index $\Gamma = -1.6 \pm 0.09$, with a $\chi^2 = 1.02$ has been obtained.
Therefore, the present data do not allow to distinguish between thermal and
non-thermal spectra.
However, if the X-ray emission were non-thermal (synchrotron and inverse
Compton) the emitting electrons would cool on timescales of few tens
of years. Therefore, the flux from the
diffuse cloud reasonably comes from a high temperature plasma.
\par
Among the discrete sources,
spectral information can be obtained only for
the radio galaxy NGC 6107 (source A$_1$), as for the other sources too few
counts have been accumulated. We assumed thermal emission, originating from
the extended corona, and again fixed N$_H$ to the galactic value. The $\chi^2$
contour plot gives a minimum at a lower temperature and higher
metallicity compared to the diffuse component discussed above (Fig. 6c). With
${\cal {M}} = 0.5$ we get best fit values $kT=1.2 \pm 0.2$ and $A=2.4 \pm 0.7
\times 10^{-4}$ photons s$^{-1}$ keV$^{-1}$, (reduced $\chi^2$=0.8, 11 d.o.f)
and a total luminosity $L_X = 1.2 \times 10^{42} h^{-2}_{50}$
erg s$^{-1}$. Fits with
larger values of the metallicity only affect these values of
temperature and luminosity within the statistical errors. The deduced energy
flux is consistent with the count rate detected in the HRI, with
one HRI count s$^{-1}$ corresponding to
$4.9 \times 10^{-11}$ erg s$^{-1}$ cm$^{-2}$.

Concerning the source B$_1$, coincident with 1615+35, we have have computed
the luminosity from the HRI photon flux assuming a power law spectrum,
as it seems reasonable from its pointlike structure (see below Sec. 5.2).
With a photon spectral index $\approx -1.5$ we get L$_X =3.5 \times
10^{41} h^{-2}_{50}$ erg s$^{-1}$. With these
parameters in the PSPC instrument a count rate
$\approx 6 \times 10^{-3}$ s$^{-1}$ is expected. This value is about half of
the photon flux detected, consistent with the fact that in the PSPC the two
sources B$_1$ and B$_2$ cannot be separated.
\blankline\noindent
\blankline\noindent
{\it 3.4. B2 1621+38 - HRI Image}
\par\noindent
The central part of the HRI field is given in Fig. 7, superimposed onto the
optical image. The 5 strongest sources detected by the HRI are listed in Table
4, where the positions, count rates and identification are presented.
Source B is identified with a red stellar object of $\sim$19$^m$ on the
red PSS plate, source C
can be identified with a star of magnitude $m_v = 9.9$ (G.S.C. cat. 3062/1869).
Source D has no optical counterpart, while source E coincides with the
radio galaxy 1622+38 (see Sect. 2).

In Fig. 8a we present the HRI image of 1621+38 (source A in tab. 4). The radial
profile of the X-ray surface brightness, obtained
as for NGC 6107 (Sect. 3.1), shows that this source
is extended ($\approx 40^{\prime \prime}$, corresponding
to a size of $\approx 35$ Kpc), with basically symmetric structure.
A possible contribution of a pointlike component is estimated to be
of $\simless$ 50\% of the total flux.
It is plausible that the X-ray emission is related to a gaseous
atmosphere of hot gas associated with the galaxy.
A hydrostatic isothermal model (Eq. 1) was fit to the observed profile,
to derive the parameters
of the gas distribution (see Sect. 4.1).

The VLA radio image at 8.4
GHz is given in Fig. 8b. The structure consists of a core, with two opposite
symmetric jets, extended 10$^{\prime \prime}$, i.e. well imbedded within the
optical and X-ray structure. The jets bend to east, to form a tail of about
1$^{\prime}$, characterized by a prominent helical shape.

To estimate the total luminosity we assume that the hot plasma has physical
parameters similar to those of the previously discussed corona of NCG 6107.
Assuming $kT=1.2$ keV and ${\cal {M}} =0.5$, 1 HRI count s$^{-1}$ corresponds
to an energy flux of $2.8 \times 10^{-11}$ erg s$^{-1}$ cm$^{-2}$, and the
luminosity is $L_X = 2 \times 10^{42} h^{-2}_{50}$ erg  s$^{-1}$.

\blankline\noindent

\centerline{\bff 4. Parameters from the X-ray  and radio data}
\blankline\noindent
{\it 4.1. Pressure in the X-ray clouds}
\par\noindent
The gas associated with the extended X-ray emission in clusters and galactic
hot coronae is expected to be distributed according to a hydrostatic
isothermal model (see Eq. 1). From the knowledge of the core radius,
$\beta$ and the X-ray central surface brightness,
it is possible to derive the central  density of the X-ray emitting gas. For
this purpose we assume that the X-ray emissivity is that expected from a an
optically thin plasma associated with an isothermal gas (Raymond and Smith
1977).
Assuming spherical symmetry,
the profiles of the gas density and of the thermal pressure are given by:
$$
n(x)={{n_c} \over {(1 + x^2)^{3 \beta/2}}}\,,\;\;\;\;\;\;\;\;\;\;
P(x) = 2 n(x) k T
\,,
\eqno(2)
$$
\noindent
where $k$ is the Boltzmann constant,
$x=r/r_c$ is the radial coordinate in units of the core radius, and
$n_c$ is the central density:
$$
n_c =\sqrt {L_X \over 4 \pi r_c^3 \Lambda(T) \Sigma_1(x_b)} \,,
\;\;\;\;\;\;\;\;\;\;\;
\Sigma_1(x_b) = \int^{x_b}_0 {x^2 \over (1+x^2)^{3 \beta}} {\hbox {d}} x \;,
\eqno(3)
$$
\noindent
with $x_b$ the maximum radius for the model fitting.

The parameters of the intergalactic medium in the group of 1615+35, as well as
those of the gaseous galactic atmospheres of 1621+38 and NGC 6107 are given in
Table 5.

\blankline\noindent
{\it 4.2. Physical parameters from radio data}
\par\noindent
{}From the radio data it is possible to obtain the value of the minimum energy
density $u_{min}$ within the radio emitting regions, corresponding
to the case of energy equally distributed between
radiating particles and magnetic field (equipartition).

The computation of minimum energy density, and consequently of internal
minimum pressure, was performed using standard formulae and assumptions
(Pacholczyk 1970), namely equal energy density in relativistic electrons
and heavy particles (k=1), the same volume occupied by the magnetic field
and the radiating electrons (filling factor $\Phi$=1), a radio spectrum
extending from 10 MHz to 10 GHz.

The values of internal pressure were computed at different locations along the
tail of 1615+35, in the jets and the tail of 1621+38 and in the outer region
of NGC 6107, using the radio maps at our disposal (see Tab. 2).
In Table 6, we give for each source the equipartition
pressure as a function of the distance from the radio nucleus.

\blankline\noindent
\blankline\noindent
\centerline{\bff 5. Discussion}
\blankline\noindent
 {\it 5.1. Hot gas in poor groups}
\par\noindent
\par\noindent
An important observational result of this paper is the detection of
extended diffuse X-ray emission in the group of 1615+35. This emission
appears to be of thermal origin and is related to the intergalactic
medium in the group. It is roughly centered between the two galaxies NGC
6109 and NGC 6107, and is elongated in N-S direction, with a maximum
size of $\approx$ 800 kpc. The total X-ray luminosity,
gas temperature and metallicity are consistent with recent results
on other groups (Price {\etal} 1991, Mulchaey {\etal} 1993), and are
lower than those for rich clusters (Edge \& Stewart, 1991). Also, the
properties of this group are consistent with those of compact galaxy
groups (Ponman \& Bertram 1993, Ponman{\etal} 1994, Ponman 1994).

With the present data, we confirm the existence of a substantial intracluster
medium in this group of galaxies, which is consistent
with the the formation of the tailed structure of 1615+35.
As suggested by Ulrich (1978), the galaxy associated with this tailed source
moves at a moderate speed through the intracluster medium ($\Delta$V = 588
km/s), therefore the tailed structure must be caused by the medium's high
density and temperature.

According to the optical data of this field (Ulrich 1978, Ekers {\etal} 1978),
the distribution of the galaxy velocities seems to indicate either the
existence of a single cluster with a rather peculiar velocity dispersion of
galaxies, or two subclusters (A and B in Fig. 1c of Ulrich 1978), of
$\sim$2 Mpc diameter, with velocities in the range 8500-9800 km/s (subcluster
A) and 9800-10600 km/s (subcluster B) respectively. The richer and more
extended (A) is centered $\sim$5' north of the radio galaxy 1615+35, and
includes both this galaxy and NGC 6107. The present X-ray data indicate the
existence of intergalactic gas in the region surrounding the two radio
galaxies NGC 6107 and NGC 6109, which, according to Ulrich (1978), are located
in a peripheral region of subcluster A. X-ray data, therefore, argue in favour
of the presence of one single cluster, centered between the two radiogalaxies.
This is also consistent with the elongated shape of the diffuse X-ray emission
being in the same direction of the galaxy distribution.

\blankline\noindent
{\it 5.2. X-ray emission from individual galaxies}
\par\noindent
\par\noindent
We have detected X-ray extended emission associated with the galaxies NGC 6107
and NGC 6137 (1621+38). These emissions are likely to originate from hot
gaseous coronae, which were discovered to be common features in early type
galaxies (Forman {\etal} 1985).

The temperature of the corona in NGC 6107 (see
sect. 3.3), is found to be fully consistent with the range of values indicated
by Forman {\etal} (1985). We then assumed for the temperature of the hot gas
associated with 1621+38 the same value as for NGC 6107.

The gas density in the coronae is of the order of 10$^{-2}$ cm$^{-3}$ (see
Table 5), consistent with values found in the literature. The typical coronal
core radius is of a few arcseconds, corresponding to a few kpc, therefore among
the smallest values found so far. Unlike 1621+38 and NGC 6107, the galaxy
1615+35 (NGC 6109) shows only an unresolved X-ray emission, which probably
originates in the galactic nucleus. As stated in sects 3.1 and 3.4, it is
possible that in 1621+38 and NGC 6107 as well the X-ray emission is partly
contributed by a pointlike nuclear component.

The presence of both resolved (thermal) and unresolved X-ray emission has been
found to be typical in a radio galaxy, although the relative strength and size
of the resolved component varies between objects (Worral \& Birkinshaw
1994a,b). The luminosity of the unresolved X-ray component correlates
well with the core radio emission, indicating that it may be
dominated by non-thermal emission associated with an inner
radio jet (Worral \& Birkinshaw 1994a,b). We report this correlation in
Fig. 9, where we have added our data (squares) to those of Worral
\& Birkinshaw (circles). In this figure we have used, for the
X-ray luminosities of 1621+38 and NGC 6107, the upper limits corresponding to
50\% and 30\% of the total luminosity, respectively. The present data are
consistent with those of Worral and Birkinshaw (1994a,b), confirming that the
unresolved X-ray emission is likely to be of non-thermal origin.

\blankline\noindent
{\it 5.3. Radio source confinement}
\par\noindent
\par\noindent
We present in Fig. 10a the superposition of the X-ray field from the PSPC of
1615+35 onto the radio emission. In Fig.
10b, we give the radio-X-ray overlay of 1621+38.
While the tail of 1615+35 is expected to be confined by the intergalactic
medium, the radiosources associated with NGC 6107 and 1621+38, are likely to
be confined by the X-ray corona associated with the galaxies.

A comparison of the pressure of the confining medium with the minimum internal
radio pressure is shown in Figs 11a, 11b and 11c, respectively for the
intergalactic gas of the 1615+35 group, and for the galaxies NGC 6107 and
1621+38. Fig. 11a shows that the beginning of the tail is close to the pressure
equilibrium with the ambient gas, while in the further points there is a larger
apparent imbalance towards the external pressure, consistent with earlier
findings (Killeen {\etal} 1988, Feretti {\etal} 1992). The same trend is
detected in NGC 6107 and in 1621+38 (Fig. 11c), although in the last case the
imbalance is not as large.

We note that the comparison can be biased by projection effects. However, these
cannot fully account for the imbalance, since this would imply unusually
extended radio structures.
It is likely that the difference is due to the numerous assumptions used for
the
calculation of equipartition parameters. For a given radio luminosity,
the non-thermal pressure within radio lobes can be raised to equal the X-ray
pressure in different ways:
i) the filling factor $\Phi $ is lowered from 1 to 0.02 - 0.06, ii)
the energy ratio k between relativistic protons and electrons
is 20 - 60, iii) electrons significantly radiate
below the adopted low-frequency cutoff of 10 MHz,
iv) equipartition conditions do not hold,
v) thermal plasma is present within the radio lobes.

Arguments against the last hypothesis are the recent data
of NGC1275 (B\"ohringer {\etal} 1993) and Cyg A (Carilli {\etal} 1994)
showing  a deficiency of X-ray surface brightness at the location
of the radio lobes.

We note that amongst our objects the radio sources associated with galaxies
presenting hot coronae have small sizes, while the source with the more
extended radio structure is pointlike at X-ray energies.
\blankline\noindent
{\it 5.4. Implications on the dark matter}
\par\noindent
{}From the X-ray luminosity we can evaluate the mass of the extended
diffuse component and of the hot coronae surrounding NGC 6107 and
1621+38. This value of the mass can be compared with the one deduced
assuming that the gas is in hydrostatic equilibrium inside a
gravitational well.
\par\noindent
{\it Diffuse component}. From the data obtained in Sec. 4.1, we can
deduce the
mass of the extended diffuse component which emits the detected X-ray
flux. Taking into account the uncertainties of the model fit, we obtain
within a core radius the following value:
$$
M_{gas}/M_{\odot} \approx 0.7 - 1.7 \times 10^{12} h^{-2.5}_{50}
$$
\noindent
As the diffuse nebula has been assumed in hydrostatic equilibrium within
a potential well, the following condition must hold:
$$
{{{\hbox {d}} P} \over {{\hbox {d}} x}}  = - m_H r_c n_h g\,,\;\;\;\;\;\;\;
g = 4 \pi G r_c m_H {{n_{h,c}} \over {x^2}} \Sigma_1(x)
\eqno(4)
$$
\noindent
where $m_H$ is the proton mass, $G$ the gravitational constant, and
$\Sigma_1(x)$ is defined in Eq. 3. In this estimate we have neglected
the mass of the galaxies and we have assumed for the gas density $n_h$
the profile given in Eq. 2. The central density thus results:
$$
n_{h,c} = \eta {k T \over 2 \pi G r_c^2 m_H^2 \Sigma_2}\,,\;\;\;\;\;\;\;\;
\;\;
\Sigma_2 = \int^{1}_0 {dx \over x^2 (1+x^2)^{3 \beta /2}} \int^{x}_0
{ x^{\prime 2} dx^{\prime} \over (1+x^{\prime 2})^{3 \beta /2}}
\,,
\eqno(5)
$$
\noindent
where $\eta=1-2^{-3 \beta /2}$. This relation provides $n_{h,c} \approx
1 - 2 \times 10^{-2}$ cm$^{-3}$ ($\gg n_c$ reported in Tab. 5), and the
total binding mass of the diffuse matter inside a core radius
consistent with the assumption of hydrostatic equilibrium is:
$$
M_{grav}/M_{\odot} \approx 1.7 - 3.6 \times 10^{13} h^{-1}_{50}
$$
\noindent
We see that the values of the total mass of the X-ray emitting gas,
$M_{gas}$, and of the mass implied by the hydrostatic equilibrium
condition, $M_{grav}$, are quite affected by the uncertainties on the
gas parameters, but their ratio is not. The disagreement by
a factor of $\approx 20$ between  $M_{grav}$ and $M_{gas}$ can be
explained by postulating that most of the nebular mass is dark,
confirming previous  results by Mulchaey {\etal} (1993),
Ponman \& Bertram (1993) and Ponman {\etal} (1994).
\par\noindent
{\it NGC 6107, 1621+38}. Fixing for both galaxies $kT=1.2$ and a radius
of $20^{\prime \prime}$, we find the following masses for the X-ray
emitting plasma in the two haloes:
$$ (M_{gas}/M_{\odot})_{NGC 6107} \approx 4 - 6 \times 10^{9}
h^{-2.5}_{50}\;\;\;\;\;\;\;\;\;
(M_{gas}/M_{\odot})_{1621+38}   \approx 3 - 5 \times 10^{9} h^{-2.5}_{50}
$$

It is interesting now to compare the luminosity of the galaxies with the mass
necessary to keep the galactic haloes in hydrostatic equilibrium in the
gravitational well.
The binding mass within a radius $r$ in the case of a spherically symmetric
isothermal corona is (see e.g. Forman {\etal} 1985):

$$
M_{grav} /M_{\odot}    \approx 1.05 \times 10^{11} {\beta}
\left [ {T \over {10^7 {\hbox {K}}}} \right ]\,
\left [ {{r_c} \over {1 {\hbox {kpc}}}} \right ]\, \left [{r \over {r_c}}
\right ]
\,
\eqno(6)
$$
\noindent
The binding masses of the two galaxies, both of the order of $\approx 10^{12}
M_{\odot}$, can be compared to their optical luminosities.
The photometric data provide a blue absolute magnitude
$M_B=-21.42$ for NGC 6107 and $M_B=-22.65$ for 1621+38.
Assuming in Eq. 6 for both
galaxies the same values of temperature and radius as above,
we get:
$$
(M_{grav}/L_B)_{NCG6107} \approx 25 - 35 h_{50},\;\;\;\;\;\;\;\;\;
(M_{grav}/L_B)_{1621+38} \approx 8 - 10 h_{50}
$$
We note that the assumed radius has to be considered a lower
limit to the extent of the halo.
The values of $M_{grav}/L_B$ are similar to those obtained by Forman {\etal}
(1985) for a sample of early-type galaxies. As for this kind of objects
$M_{stellar}/L_B \lapp 10 h_{50}$ is expected, a dark massive halo must be
present, especially in the galaxy NGC 6107.
\blankline\noindent
\blankline\noindent
\centerline{\bff 6. Conclusions}
\blankline\noindent
\par
We have studied the X-ray emission of two groups of galaxies, containing
3  extended radio galaxies, of which two show tailed radio structure.
The conclusions of this paper can be summarized as follows.
\par\noindent
1. The group of 1615+35 is permeated by diffuse emission of about 20'
in size, which indicates the presence of hot intergalactic gas.
Its X-ray spectrum is fit by a thermal
Raymond-Smith model, with temperature $\sim$ 2.1 keV and metallicity $\sim$
0.1.
These gas parameters are typical for groups of galaxies.
The spatial distribution of the diffuse emission in 1615+35 is
consistent with a hydrostatic isothermal model.
\par\noindent
2. The X-ray source associated with the radio galaxy 1615+35 is pointlike.
Therefore it is likely originating in the galactic nucleus.
\par\noindent
3. The X-ray emission from the galaxy NGC 6107 is extended and
likely associated with a hot gaseous corona, with a gas
distribution consistent with the hydrostatic isothermal model.
We estimate that a possible pointlike component can only contribute
less than $\sim$ 30\% to the detected count rate.
The spectrum can be fit by a
thermal model, with  temperature $\sim$ 1.2 keV and metallicity $\sim$ 0.5.
\par\noindent
4. The radio galaxy 1621+38 shows extended X-ray emission,
similar to NGC 6107, associated with a hot galactic atmosphere.
The contribution from a pointlike nuclear source is, in this case,
less than $\sim$ 50\%.
\par\noindent
5. The X-ray luminosities (or upper limits) of the pointlike components
correlates with the powers of the radio core, consistent with the results
of Worral and Birkinshaw (1994a,b). This favours the X-ray emission to
be of non-thermal origin associated with the inner radio jets.
\par\noindent
6. The internal equipartition pressure in the outer radio tail
of 1615+35 is lower
than the pressure of the external intergalactic medium. The same trend is found
for the outer regions of the  radio galaxies NGC 6107
and 1621+38, on the smaller  galactic scale.
This implies that  either the assumptions made in the computation  of the
internal pressure are not valid, or that there is a real deviation from
equipartition conditions.
\par\noindent
7. The  total mass of the X-ray emitting intracluster gas in the group of
1615+35 is much lower than that deduced by the hydrostatic equilibrium
assumption. This can be used as evidence for the existence of significant dark
matter associated with this group. Similarly, massive dark halos must be
present in the galaxies NGC 6107 and 1621+38.
\blankline\noindent
\blankline\noindent
\centerline{\bff Acknowledgements}
\blankline\noindent
We wish to thank R. Morbidelli and M. Lattanzi for providing the optical PSS
maps, and A. Volpicelli for software assistance. Thanks are due to R.McMahon
for providing the APM finding charts, to B. Clark for scheduling the VLA
observation of 1622+38, and to H. B\"ohringer for the programs of spatial
analysis and for helpful discussions. We are indebted to N. Primavera for the
reproduction of the figures.
\par\noindent
This research has made use of the NASA/IPAC Extragalactic Data Base (NED) which
is operated by the Jet Propulsion Laboratory, California Institute of
Technology, under contract with the National Aeronautics and Space
Administration.
\vfill\eject
\skuno
\singlespace
\centerline{\bff References}
\blankline\noindent
\skuno
\parref
Allen, C.W., 1973, Astrophysical Quantities, The Athlone Press, Univ. of London
\parref
B\"ohringer, H., Voges, W., Fabian, A.C., Edge, A.C., Neumann, D.M., 1993,
MNRAS 264, L25
\parref
Burns, J.O., 1986, In: Henriksen, R.N. (ed.), Jets from Stars and Galaxies,
Can. J. Phys. 64, p. 373
\parref
Burns, J.O., Hanisch, R. J., White, R.A., Nelson, E.R., Morrisette, K.A.,
Ward Moody, J., 1987, AJ 94, 587
\parref
Carilli, C.L., Perley, R.A., Harris, D.E., 1994, MNRAS, in press
\parref
Cavaliere, A., Fusco-Femiano, R., 1976, A\&A 49, 137
\parref
Cavaliere, A., Fusco-Femiano, R., 1981, A\&A 100, 194
\parref
Colla, G., Fanti, C., Fanti, R., Gioia, I., Lari, C., Lequeux, J., Lucas,
R., Ulrich, M.-H., 1975, A\&AS 20, 1
\parref
Dickey, J.M., Lockman, F.J., 1990, ARA\&A 28, 215
\parref
Edge, A.C., Stewart, G.C., 1991, MNRAS 252, 428
\parref
Ekers, R.D., Fanti, R., Lari, C., Ulrich, M-H: 1978, A\&A 69, 253
\parref
Fanti, R.: 1984, In: Mardirossian F., Giuricin, G., Mezzetti M. (Eds.),
 Clusters and Groups of Galaxies, Reidel P.C., Dordrecht, p.185
\parref
Fanti, C., Fanti, R., De Ruiter, H.R., Parma, P., 1987, A\&AS 69, 57
\parref
Feretti, L., Spazzoli, O., Gioia, I.M., Giovannini, G., Gregorini, L.: 1990,
A\&A 233, 325
\parref
Feretti, L., Perola, G.C., Fanti, R: 1992, A\&A 265, 9.
\parref
Forman, W., Jones, C., Tucker, W, 1985, ApJ 293, 102
\parref
Gregory, P.C., Condon, J.J., 1990, ApJS 75, 1011
\parref
Harris, D. E., Forman, W., Gioia, I. M., et al., 1990,
The Einstein Observatory catalog of IPC X ray sources,
Smith. Astr. Obs., Cambridge (USA), Volume 6E
\parref
Killeen, N., Bicknell, G.V., Ekers, R.D.: 1988, ApJ 325, 180
\parref
McMahon, R.G., 1991, in The Space Distribution of Quasars, Ed. D. Crampton,
ASP Conference Series, Vol 21, p 129
\parref
Morganti, R., Fanti, R., Gioia, I.M., Harris, D.E., Parma, P., de Ruiter, H.,
1988, Astron. Astrophys.,  189, 11
\parref
Mulchaey, J.S., Davis, D.S., Mushotzky, R.F., Burstein, D., 1993, ApJ 404, L9
\parref
Noonan, T.W., 1973, AJ 78, 26
\parref
Pacholczyk, A.G., 1970, Radio Astrophysics, Freeman and Co., San Francisco.
\parref
Pfeffermann, E., Briel, U. G., Hippmann, H.,
et al., 1987, Proc. Society for Photo-Optical
Instrumentation Engineers (SPIE) 733, 519
\parref
Ponman, T.J., Bertam, D., 1993, Nat 363, 51
\parref
Ponman, T.J., Allan, D.J., Jones, L.R., Merrifield, M., McHardy I.M.,
Lehto, H.J., Luppino, G.A., 1994, Nat 369, 462
\parref
Ponman, T., 1994, Paper presented at the Rencontres de Moriond, Clusters
of galaxies, in press
\parref
Price, R., Burns, J.O., Duric, N., Newberry, M.V., 1991, AJ 102, 14
\parref
Raymond, J.C., Smith, B.W., 1977, ApJS 35, 419
\parref
Rothenflug, R., Arnaud, M., 1985, A\&A 144, 431
\parref
Sarazin, C.L., 1986, Rev. Mod. Phys. 58, 1
\parref
Tr\"umper, J., 1983, Adv. Space Res. 2, 241
\parref
Ulrich, M.-H., 1978, ApJ 221, 422
\parref
Worral, D.M., Birkinshaw, M., 1994a, IAU Symp. n. 159, on
``Multi-wavelength continuum emission of AGN", Eds. Courvoisier \& Blecha,
p. 385
\parref
Worral, D.M., Birkinshaw, M., 1994b, ApJ in press
\parref
Zimmermann, H.U., Belloni, T., Izzo, C., Kahabka, P., Schwentker, O., 1993,
EXSAS User's Guide, MPE Report 244
\parref
Zwicky, F., Herzog, E., 1966, Catalogue of Galaxies and Clusters of
Galaxies, Vol. III, Cal. Inst. Technology
\vfill\eject
\centerline {\bff Figure caption}
\par\noindent
{\bf Fig. 1} - Isocontour HRI map of 1615+35 smoothed with a gaussian of
12$^{\prime \prime}$ (FWHM), overimposed onto the grey-scale blue image of the
same field. Contours are at 17, 24, 34, 51, 68 and 84 \% of the peak
brightness. Sources are labelled according to Table 1. In this and the
following figures, coordinates are equatorial J2000.
\par\noindent
{\bf Fig. 2} - {\bf a}: HRI map of the sources B$_1$ and B$_2$, smoothed with a
gaussian of 12$^{\prime\prime}$ (FWHM). Contours are at 20, 31, 42, 52, 63 and
84\% of the peak brightness of 0.48 counts arcsec$^{-2}$. The first contour
corresponds to a level 3 $\sigma$ above the background. {\bf b}: VLA radio  map
at 1.4 GHz of the source 1615+35. Resolution is
4.6$^{\prime\prime}\times3.4^{\prime\prime}$ (@-13$^{\circ})$. Contours are at
-0.6, 0.6, 1.2, 1.8, 3.0, 4.2, 5.4, 7, 10, 15 and 20 mJy/beam. The radio core,
indicated by an arrow, coincides with the westernmost X-ray peak.
\par\noindent
{\bf Fig. 3} - {\bf a}: HRI map of the source A$_1$, smoothed with a gaussian
of 12$^{\prime\prime}$ (FWHM). Contours are at 15, 22, 30, 37, 44 and 60 \% of
the peak brightness of 0.69 counts arcsec$^{-2}$. The first contour corresponds
to a level 3 $\sigma$ above the background.  {\bf b}: VLA radio map at 1.4 GHZ
of the source NGC 6107, identified with the X-ray source A$_1$. Resolution is
13$^{\prime\prime}\times12^{\prime\prime}$ (@-6$^{\circ})$. Contours are at
-0.4, 0.4, 0.6, 1.2, 1.8, 3.0, 4.2, 5.4, 7, 10, 15 and 20 mJy/beam.
\par\noindent
{\bf Fig. 4} -  {\bf a}: PSPC map of the field of 1615+35 smoothed with a
gaussian of 50$^{\prime \prime}$ (FWHM). Contours are at 13, 25, 38, 50, 63, 76
and 95 \% of the peak brightness of 0.049 counts arcsec$^{-2}$. The first
contour corresponds to a level 3 $\sigma$ above the background.  {\bf b}: PSPC
map of the field of 1615+35, smoothed with gaussians of increasing width with
decreasing count rate up to 4.5$^{\prime}$, to enhance the low brightness
diffuse structure. Contours are at 2, 5, 7, 9, 12, 24, 36, 60 and 83 \% of the
peak brightness of 0.038 counts arcsec$^{-2}$. The first contour corresponds to
a level 3 $\sigma$ above the background.
\par\noindent
{\bf Fig. 5} -  X-ray radial profile of the diffuse emission in the 1615+35
cluster, corrected for vignetting and after removal of point sources.
The radius is in arcsec, the surface brightness in counts s$^{-1}$
arcsec $^{-1}$. The solid line
represents the best fit model. The background is fit.
\par\noindent
{\bf Fig. 6} - {\bf a}: Confidence levels at 1, 2 and 3 $\sigma$ of metallicity
${\cal {M}}$ versus temperature $kT$ for the diffuse source in the 1615+35
cluster assuming a thermal Raymond-Smith spectrum, with fixed N$_H$. Contours
are for {\bf b}: Spectrum od the diffuse source in 1615+35, with the solid line
representing the best fit. {\bf c}: Same as panel a, for the X-ray source
associated with the galaxy NGC 6107.
\par\noindent
{\bf Fig. 7} -  Isocontour HRI map of 1621+38 smoothed with a gaussian of
7$^{\prime \prime}$ (FWHM), overimposed onto the grey-scale blue image of the
same field. Contours are at 16, 22, 33, 43 and 54 \% of the peak brightness.
Source A is the radio galaxy 1621+38, while the other sources of Table 2 are
not in the figure.
\par\noindent
{\bf Fig. 8} -  {\bf a}: HRI map of the source 1621+38 smoothed with a gaussian
of 7$^{\prime \prime}$ (FWHM). Contours are at 13, 22, 33, 43, 54, 65, 76 and
87 \% of the peak brightness of 9.2 counts arcsec$^{-2}$. The first contour
corresponds to a level 3 $\sigma$ above the background.  {\bf b}: VLA radio map
at 8.4 GHz of 1621+38. Resolution is
2.2$^{\prime\prime}\times2.0^{\prime\prime}$ (@6$^{\circ})$. Contours are at
-0.12, 0.12, 0.25, 0.5, 1, 1.5, 2.5 5, 10 and 25 mJy/beam.
\par\noindent
{\bf Fig. 9} -  Plot of the spectral X-ray luminosity at 1 keV
of the pointlike component versus the 5 GHz radio core luminosity
for our galaxies (squares) and those analyzed by Birkinshaw and Worral
(1994a,b) (circles).  For 1621+35 only the upper
limit of the X-ray luminosity is reported, while for NCG 6107 the upper limit
is for both the X-ray and radio luminosity.
\par\noindent
{\bf Fig. 10} -  {\bf a}: Contour map of the radio emission at 0.6 GHz
of the 1615+35 field overimposed onto the gray-scale X-ray image.
Resolution of radio map is 51$^{\prime\prime}\times30^{\prime\prime}$
(@0$^{\circ})$. Levels are at 2, 6, 15, 27, 50 and 100 mJy/beam.
Discrete X-ray
sources are identified with the tailed radio galaxy 1615+35, and with NGC 6107.
It is well visible the diffuse X-ray emission permeating the cluster.
{\bf b}: Contour map of the radio emission at 8.4 GHZ of 1621+38, superimposed
onto the grey-scale X-ray image. Resolution of radio map is
2.2$^{\prime\prime}\times2.0^{\prime\prime}$
(@6$^{\circ})$. Levels are 0.25, 0.5, 1, 2, 3, 5 and 10
mJy/beam.
\par\noindent
{\bf Fig. 11} - Plot of the equipartition radio pressure (dots) and of
the allowed range for the
thermal pressure of the X-ray emitting gas (dashed)
versus the radial distance for
{\bf a:}  the 1615+35 group, {\bf b:}  NGC 6107, {\bf c:} 1621+38.
The distance is in arcsec, the pressure in dyn cm$^{-2}$.
\vfill\eject
 TABLE 1 - Observational details
\skuno
\settabs
\+sourcesource & Instrinstr & radecradecradecradecradecradec &
datedatedatedatedate & exposure \cr
\+ Name   &  Instr    & RA \ \ \ \ (J2000)\ \ \ \ DEC & Date & Exposure \cr
\+        &           & \ h \ m \ s \ \ \ \ \ \ \ \ \ $\circ$\ \
$\prime$ \ $\prime\prime$ & & sec \cr
\skuno
\+ 1615+35  & HRI  &  16\ 17\ 38.4\ \ \ \ \  35\ 00\ 00 & Aug 91  & 33740 \cr
\+          & PSPC &  16\ 17\ 38.4\ \ \ \ \  35\ 00\ 00 & Jan 92 /
Jan 93  &  6404 \cr
\+ 1621+38  & HRI  &  16\ 23\ 00.0\ \ \ \ \ 37\  55\ 12 &  Aug 91  & 31800 \cr
\skuno
\skuno
\skuno
TABLE 2 - Radio maps
\skuno
\settabs
\+ sourcesource & instrinstrinstr & freqfreq & timetimetime & hpbwhpbwhpbwhpbw
 & noisenoise &
refref \cr
\+ Name   & Instr  & Freq &  Obs Time & HPBW  & Noise & Ref \cr
\+        &        & MHz  &       &  $\prime\prime$\ \ \ \ ($\circ$)  &
mJy/b \cr
\skuno
\+ 1615+35  & WSRT     &  608   & 12 h  & 51$\times$30(@0) & 1.0  &
Unpublished \cr
\+          & VLA - C  &  1425  & 12 m  & 13$\times$12(@-6) &  0.15 &
Pres paper \cr
\+          & VLA - B  &  1415  & 18 m   & 4.6$\times$3.4(@-13) & 0.08 &
Unpublished\cr
\+ \cr
\+ 1621+38  & VLA - C  &  8440  &  12 m & 2.2$\times$2.0(@6)  &   0.04  &
Pres paper \cr
\+          & VLA - A  &  1440  &  14 m  & 1.4$\times$1.3(@-52) &  0.27 &
Fanti et al.1986 \cr
\+          & VLA - B  &  1465  &  34 m  & 3.5$\times$3.3(@-36)  & 0.20 &
   Parma et al. 1986 \cr

\skuno
\skuno
\skuno
 TABLE 3 - Sources in the field of 1615+35
\skuno
\settabs 10\columns
\+ Source  & \ \  RA \ \ \ \ \ \ (2000) \ \ \ \ \ \ DEC &&&&
CR$_{HRI}$ && CR$_{PSPC}$ && Id  \cr
\+ &   h\ \ \ \ m\ \ \ \ s && $\circ$ \ \ \ \ $\prime$ \ \ \ $\prime \prime $
&&  $\times 10^{-3}$ \ s$^{-1}$ && $\times 10^{-2}$ \ s$^{-1}$  \cr
\skuno
\+ A$_1$ & 16 \ 17 \ 20.2 && 34 \ 54 \ 05 &&   6.5 $\pm$ 0.6
&& 3.0 $\pm$ 0.3 && NGC 6107 \cr
\+ A$_2$ & 16 \ 17 \ 18.2 && 34 \ 54 \ 51 &&   1.5 $\pm$
0.6&& 0.28 $\pm$ 0.09 && Star \cr
\+ B$_1$ & 16 \ 17 \ 40.8 && 35 \ 00 \ 16 &&   1.9 $\pm$ 0.3
&& 1.2 $\pm$ 0.6 && NGC 6109 \cr
\+ B$_2$ & 16 \ 17 \ 43.3 && 35 \ 00 \ 05 &&   1.1 $\pm$ 0.3
\ \ {\raise 4pt \hbox {$\}$}}&&  && Blue obj \cr
\+ C & 16 \ 17 \ 36.5 && 35 \ 04 \ 49 &&  2.1 $\pm$ 0.4
&&  2.1 $\pm$ 0.3  && Blue obj \cr
\+ D     & 16 \ 17 \ 25.6 && 35 \ 07 \ 25 &&  1.4 $\pm$ 0.4
&&  0.8 $\pm$ 0.2  && Blue obj \cr
\vfill\eject
 TABLE 4 - Sources in the field of 1621+38. Positions are corrected for
the positional offset mentioned in the text.
\skuno
\settabs 10\columns
\+ Source  & \ \  RA \ \ \ \ \ \ (2000) \ \ \ \ \ \ DEC &&&&
CR$_{HRI}$ && Id  \cr
\+ &   h\ \ \ \ m\ \ \ \ s && $\circ$ \ \ \ \ $\prime$ \ \ \ $\prime \prime $
&&  $\times 10^{-3}$ \ s$^{-1}$ \cr
\skuno
\+ A  & 16 \ 23 \ 03.1 && 37 \ 55 \ 21 && 17 $\pm$ 1
&& NGC 6137 \cr
\+ B  & 16 \ 23 \ 34.3 && 37 \ 49 \ 51 && 2.8 $\pm$ 0.7
&& Red obj \cr
\+ C  & 16 \ 22 \ 56.4 && 38 \ 04 \ 26 && 3.8 $\pm$ 0.7
&& Star \cr
\+ D  & 16 \ 24 \ 03.3 && 38 \ 01 \ 16 && 4.6 $\pm$ 0.7
&& - \cr
\+ E  & 16 \ 24 \ 16.7 && 37 \ 49 \ 48 && 7.1 $\pm$ 0.8
&& Radio Gal \cr
\skuno
\skuno
TABLE 5 - Parameters of the X-ray gas
\skuno
\settabs
\+ Sourcesourcesourcesourc & densitydensitydensdens    & radiusradiusradius
& radiusradiusradius & temptemptemp \cr
\+ Name       &   n$_c$ & $\beta$ & r$_c$  & Temp \cr
\+            &   $\times h^{1/2}_{50}$cm$^{-3}$        &  &  $
\prime\prime$ & keV \cr
\skuno
\+ 1615+35(Diffuse)    &  (6.1 $\pm$ 0.6)$\times$10$^{-4}$   &
  0.56 $\pm$ 0.07  &   375 $\pm$ 85   & 2.1 $\pm$ 0.6  \cr
\+ NGC 6107            &  (4.9 $\pm$ 0.8)$\times$10$^{-2}$   &
0.47 $\pm$ 0.07  &    4.5 $\pm$ 2.0   & 1.2 $\pm$ 0.2 \cr
\+ 1621+38             &  (5.1 $\pm$ 0.8)$\times$10$^{-2}$&
0.44 $\pm$ 0.04  &    3.5 $\pm$ 1.5  & 1.2$^*$  \cr
\skuno
* = Assumed
\skuno
\skuno
TABLE 6 - Parameters of the radio emitting regions
\skuno
\settabs
\+ Sourcesourcesourc & densitydensit    & pressure \cr
\+ Name            &   Dist. & P$_{eq}$ \cr
\+        &   $ \prime\prime$ & $\times h^{4/7}_{50}$ dyn cm$^{-2}$ \cr
\skuno
\+ 1615+35    &      64     &     1.29$\times$10$^{-12}$   \cr
\+            &     128     &     1.01$\times$10$^{-12}$   \cr
\+            &     227     &     4.72$\times$10$^{-13}$   \cr
\+            &     305     &     2.27$\times$10$^{-13}$   \cr
\+            &     420     &     1.14$\times$10$^{-13}$   \cr
\+            &     643     &     6.10$\times$10$^{-14}$   \cr
\+            &     782     &     5.97$\times$10$^{-14}$   \cr
\skuno
\+ NGC 6107   &     25       &   7.48$\times$10$^{-13}$   \cr
\skuno
\+ 1621+38    &     5(North) &    6.93$\times$10$^{-11}$   \cr
\+            &     5(South) &   5.43$\times$10$^{-11}$   \cr
\+            &     20       &   9.98$\times$10$^{-12}$   \cr
\+            &     35       &   2.95$\times$10$^{-12}$   \cr
\bye